# Study and Analysis of MAC/IPAD Lab Configuration

Ayman Noor
Department of Computer Science
College of Computer Science and Engineering
Taibah University, Madinah
Saudi Arabia
anoor@taibahu.edu.sa

**Abstract**
This paper is about three virtualization modes: VMware™, Parallels™ and Boot Camping™. The trade off of their testing is the hardware requirements. The main question is, among the three, which is the most suitable? The answer actually varies from user to user. It depends on the user needs. Moreover, it is necessary to consider its performance, graphics, efficiency and reliability, and interoperability, and that is our major scope. In order to take the final decision in choosing one of the modes it is important to run some tests, which costs a lot in terms of money, complexity, and time consumption. Therefore, in order to overcome this trade off, most of the research has been done through online benchmarking and my own anticipation. The final solution was extracted after comparing all previously mentioned above and after rigorous testing made which will be introduced later in this document.

## 1. Introduction

### 1.1 Overview

As a start on this paper, we will try to find the differences between Parallels™, VMware Fusion™, and Boot camp™ in term of cost and performance. Then, at the end of this document after comparing the benchmark and the rigorous testing in different conditions and environments, a suggestion will come. The final recommendation for a solution is based on the investigation completed. The idea in this project is to find a cost- and performance-effective way to match Windows OS and its applications with the Apple OS and its applications on one computer. The goal is to use this investigation to build a teaching lab that can do application development for both Apple/iOS and Windows or Android. There three competing approaches to be compared are commercial software packages that support dual Mac OS and Windows operating systems on a single computing platform (Intel or Mac OS). The three competing solutions investigated were Parallels™, VMware™, and Boot camp™, with the expectation that the Computer and Information Science (CIS) department will implement in the near future. The project encompasses performance, cost and trade-off analysis that is presented later in this document.

### 1.2 Curriculum Scope

It has been experimental and a learning experience. As it entails lots of research and practical testing to take the final verdict about which mode to go with in this work, and despite the testing cost of it, the solution provided is the optimal decision.

### 1.3 Project Management Plan

The project management plan was one of the most important aspects of the project. A few issues were encountered while developing the project and its testing. One major issue of this project was the cost because of the need of two different environments of operating system. Time was another factor because of the deadline that was set on the project. Since a lot of research was put into figuring out the different specifications from the different benchmarks, testing was done simultaneously. Testing for this project requires the hardware of different specifications and loads that were put on the systems to reach some solid solution. Another time-consuming task was testing the frame rates each system runs on with the video games. This took planning, concentration, and cost to provide the favorable results that were achieved.

### 1.4 Literature Perspective

"Boot Camp™, VMware Fusion™, and Parallels™ are programs that assist users in installing Microsoft Windows XP or Windows Vista on Intel-based Macintosh computers. Due to performance benefits, the School of Architecture + Design recommend using Boot Camp™ for running Windows on a Mac." [1][12][13].

### 1.5 Boot Camp™

Apple offers a dual-system called Boot Camp™. This software is free and every new Mac lets you install and run Windows at its normal speed. The setup is simple and safe for Mac files. After the installation has completed, the option to boot up the Mac in OS X or Windows is presented. The Windows operating system will run at its normal speed without having to worry about slow performance. The downside to Boot Camp™ is that the Mac will have to re-boot each time the operating system needs to be switched. Boot Camp™ is simply a dual boot style, so if Windows is installed with Boot Camp™ the downside is that the Mac will have to reboot to use Windows, which is not the case for virtual machines. However, a good thing is that 100% of the system resources will be available. Another plus to boot camp™ is that it is a good choice for gamers. The reason is that it offers better graphic quality. [2] The Boot Camp™ program gives the ability to choose between Windows OS and Mac OS at the Mac boot menu. The interesting thing that Boot Camp™ is able to do is running Windows at its original speed without the virtualization layers that come with Parallels™ and VMware Fusions™. This could either be a positive thing or a negative thing depending on specific requirements. Boot Camp™ is able to run Windows natively on a Mac. This could be a negative thing for some users because data cannot be shared easily between Windows and the Mac OS. It's a positive thing for users who enjoy Windows applications that include video editors, CAD programs, and games. [3][12][13].

1.6 VMware Fusion™ and Parallels ™ Overview

VMware Fusion™ and Parallels™ are virtual machine programs. A virtual PC through software is created. The thing that VMware Fusion™ and Parallels™ have in common is that both allow Macintosh users to be able to run Windows applications without having to reboot the computer.  The downside to this is that both of the machine programs will not have 100% of the system resources available, and this in turn will lessen the performance. Both Parallels™ and VMware Fusion™ are simple to use and install. Both come with a useful guide that makes it simple for users to create a virtual machine. These products offer tools as well as special drivers for Windows that are able to improve the performance in a virtual environment. Things that are offered in these tools include a shared folder that can be retrieved from Windows and Mac OS X. [4]

1.7 VMware Fusion™

VMware Fusion™ is able to choose whether the Intel Mac will need to use one or both core processors. This is a good feature because the user will have the choice to give the virtual machine all of the processing power. [4] The price for VMware Fusion™ is $47.00. [5]However, if the user plays video games, then VMware™ is not the right path to go. This is because; if the user is running VMware Fusion™ while playing a game it would be like running two operating systems at once. This is bad because all of the system resources are being used, which in turn lessens the quality of the game graphics. [6]

1.8 Parallels™

Parallels™ is even simpler than VMware Fusion™ because it provides "Express Windows OS Installation Mode" which has the ability to perform all of the installation processes of either Windows XP or Vista. It is a simple process and all that needs done is to input a username as well as a Windows activation key. This is helpful because it is easy to use and very convenient. Another positive thing is the clipboard support which allows the user to copy and paste between Mac and Windows applications. The price for Parallel™ is $79.99 or $49.99 to upgrade from a previous version. [5] Out of VMware Fusion™ and Parallels™, Parallels™ is much more reliable and user friendly. [4]

## 2. Comparison of VMware Fusion™, Parallels™, and Boot Camp™

2.1 Opening and Closing

Although the two-virtualization apps do differ in speed in term of the speed with which they open, sleep, resume, and shut down, testing shows that Parallels™ is notably faster at each of those tasks. Especially in suspending and resuming. Therefore, the user needs to open and close virtual machines all day, these time savings could add up. [7] Both of these virtualization apps are stable. There are no outright crashes in either of them; however, there are some slight downsides on both of them. When it comes to VMware Fusion™, the full screen mode exiting and entering causes more redraws than in Parallels™. In addition, some apps fail in Windows when using Parallels™, as well as not being able to type passwords at the Linux login prompt; note that this did not happen in VMware Fusion™. [7]

2.2 Virtualizing Windows

Overall, both VMware Fusion™ and Parallels™ do very good jobs. They both are capable of supporting hundreds of guest operating systems. It comes in handy for users, especially when trying to run more than one version of Windows. [7] From earlier reviews, both Parallels™ and VMware Fusion™ did an excellent job in running on earlier versions of Windows. Therefore, the upcoming version of Windows, which is Windows 8, was focused on. In testing, Windows 8 Developer Preview was handled well by both applications. [7] In the traditional Windows interface the Desktop button is on the Start menu, both of these apps are able to run Windows and its predecessors. While testing, office applications ran well and without any delays. The system did not feel slow in either of the programs. Moreover, two different email applications were tested, and both worked fine. Also, browsing the internet went smoothly and the window interface was fast. [7]

### 2.3 Hardware & Software

On iMac at 3.4 GHz with 16 GB of RAM running OS X 10.8.1, tests were performed to virtualize Windows 7 Professional 64-bit. VMware Fusion™ 4.1.3, VMware Fusion™ 5.0.1, Parallels™ and Virtual Box 4.2 were used. [8] All the machines were stored on and accessed from an external Pegasus R4 Thunderbolt RAID array with four 3 TB 7200 rpm hard drives in a RAID 5 configuration. Windows via Boot Camp™ was unable to be installed on this drive, so native Windows was employed and installed on the iMac's internal 240 GB OWC SATA III SSD [8]. The machines were then configured to use four of the iMac's eight logical processors and 4 GB of RAM. The configuration doesn't completely translate between applications, but were set up to get the most out of the performance of the virtual machine over that of the host OS. For the Boot Camp tests, all benchmarks were set to use only four cores to provide a more accurate comparison. [8]

### 2.4 Literature Summary

As stated above the comparison of VMware Fusion™ and Parallels™ was completed followed by all three together for better understanding of the issue and to suggest the best solution. Different testing were used and at the end of a decision was made which has good performance and integrity-wise. Until now, Boot Camp™ has been found to be the better solution for all those who cannot compromise on performance.

### 2.5 Testing Methodology

The results of testing each configuration for three times were average as long as they were within five percent of each other. For each odd result that has been occur, we retest until we find the source that causes the irregularity then that result is discarded. Furthermore, not all tests were possible in every configuration. These exceptions were duly noted. [8]

### 2.6 Objectives of the Analysis

The purpose of this analysis is to rigorously test the Boot Camp™ via different testing methodology. Finally, the proposition will be supported strongly and will be appealing with full justification as to why Boot Camp™ is the best decision.

### 2.7 Methodology Application

The methodologies of testing and different techniques are as follows in Fig. 1:

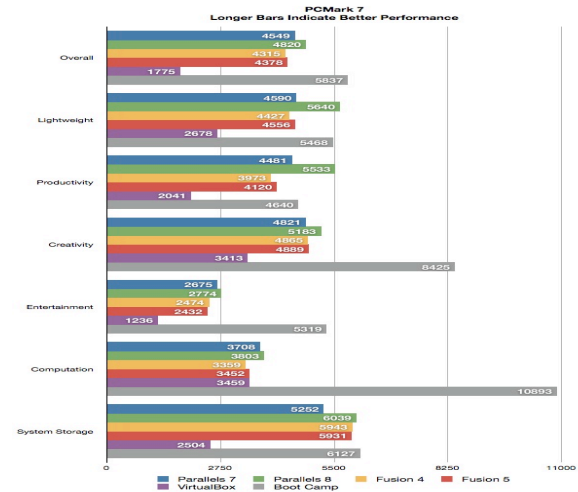

Fig. 1: PCMark7

First, it should be noted that virtual benchmarks, such as the Lightweight and Productivity tests report artificially high scores. This is primarily due to the inability of having a fair and even CPU comparison between virtual and native hardware, and due certain virtualization optimization that can inflate synthetic benchmark scores. [8] It is also important to notice the Boot Camp™ result for the Computation test. Restricting the number of CPU cores is not allowed in PCMark 7 for the user. The Boot Camp™ tests were able to use four more logical cores than the virtualized tests that were restricted to four. Boot Camp™ performance would still be superior to virtualized performance, but not significantly larger. [8] Although virtualized performance is not far behind in some categories. Boot Camp™ is the obvious winner. As previously demonstrated, Parallels™ 8 offers the best virtualization performance for the test, with Parallels™ 7 in second and Fusion 4 and 5 in third and fourth. The free VirtualBox has a lot of catching up to do in order to match the performance of the commercial software applications and it comes in 5[th]. [8]

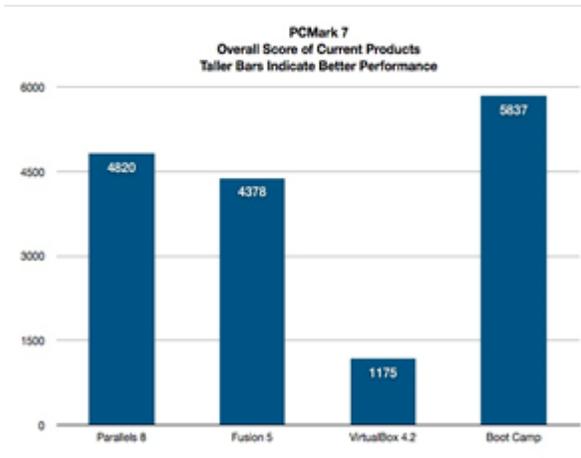

Fig. 2: PC Mark 7

As showing in Fig. 2, the virtualization software is solely focused on as compared to Boot Camp™; it is evident that Parallels™ 8 has an approximate 10% advantage over VMware Fusion™ 5, while native Boot Camp™ tacks an about 21% higher performance over Parallels™ 8. [8]

## 2.8 3DMark06

3DMark06 is a DirectX 9 gaming benchmark that stresses a system's GPU and CPU in basically the same way that a highly graphical game would. Despite the fact that it seems a little outdated at this point, it still provides one of the best procedures to test DirectX 9 performance for which both Parallels™ and VMware Fusion™ offer full support. Virtual Box has basic 3D support but would not run the 3DMark06 benchmark, Fig 3. [8]

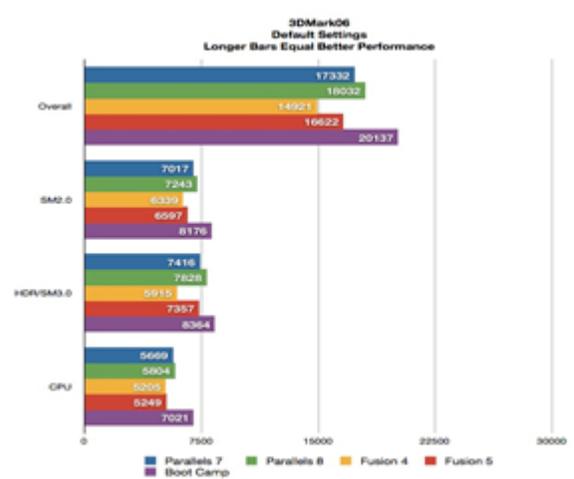

Fig. 3: 3D Mark 06

"The results show the same trend found in the PCMark 7 tests: Boot Camp™ takes first, Parallels™ 7 and 8 lead the virtualization results, and VMware Fusion™ 4 and 5 finish last." [8]

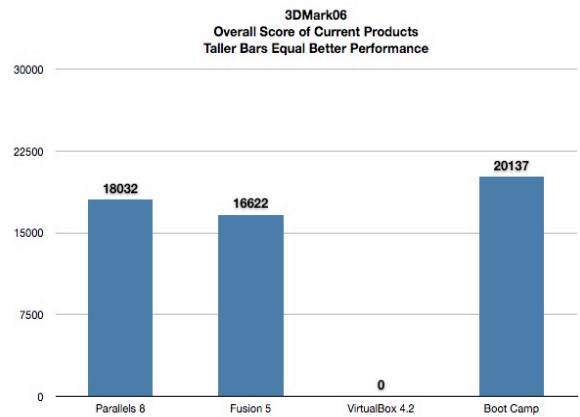

Fig. 4: 3D Marks 06

Aiming the focus on the current versions of virtualization applications, it was observed that Parallels™ 8 scores about 8.5% greater than VMware Fusion™ 5, and roughly 11.5 % lower than Boot Camp™, Fig. 4. [8]

## 2.9 3DMark Vantage

3DMark Vantage is a DirectX 10 benchmark. While the iMac's GPU supports DirectX 10 natively in Boot Camp, Parallels™ 8 is the only one that supports the multimedia API while it is virtualized. Parallels™ categorize the DirectX 10 support as "experimental," so future improvements in performance are seemingly inevitable. [8] 3DMark Vantage has numerous presets for its CPU and GPU calculations. A default setting for Performance (P) and Entry (E) tests was implemented, Fig. 5. [8]

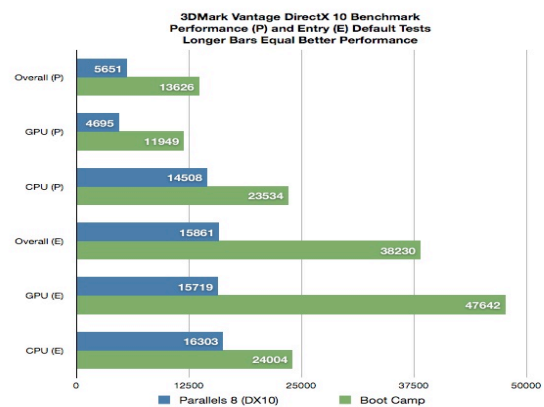

Fig. 5: 3D Mark Vantage

Test results demonstrate that DirectX 10 support does have some ways to go yet. While it does function, gamers shouldn't get too eager trying to virtualize their favorite DX10 games. [8]

## 2.10 Geekbench

Geekbench is a multi-platform tool for assessing a system's memory performance and computation. It doesn't necessarily test storage capabilities or even graphics but it does have its uses in which it scales from systems as small as a smart phone or tablet to machines using a dozen processors. [8] "Geekbench can be run in either 32- or 64-bit mode. As we were using a 64-bit version of Windows with 4 GB of RAM, we ran the 64-bit mode, Fig. 6." [8]

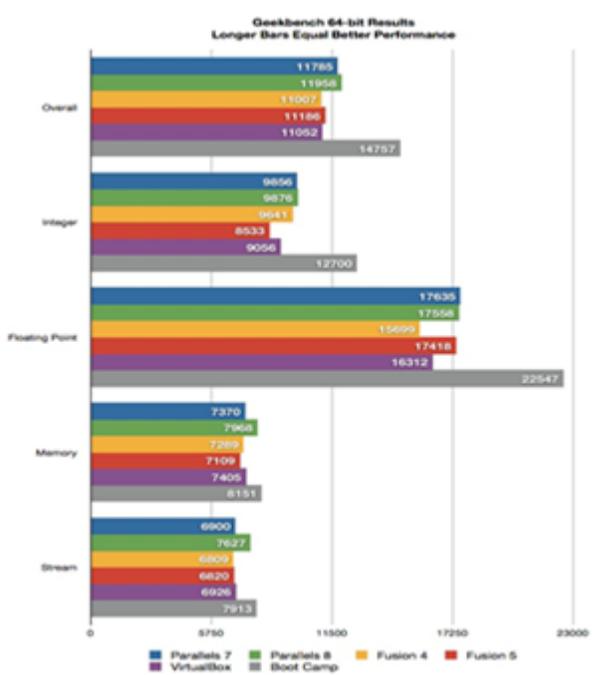

Fig. 6: Geekbench 64-bit

With no surprises, Parallels™ has a clear and distinct advantage over VMware Fusion™ in most trials, although both applications came close to native performance in both memory and stream tests. The free Virtual Box can run these tests as well and can hold its own. It may not win any categories but it does however keep within 10% of Parallels™ on most of the tests. Boot Camp™ holds a substantial lead in Integer and Floating Point calculations, Fig. 7. [8]

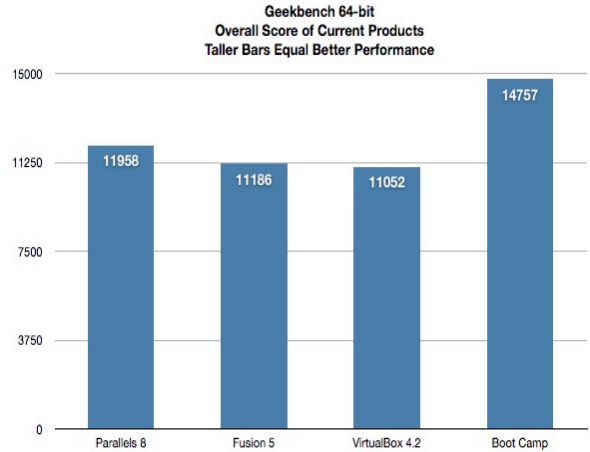

Fig. 7: Geekbench 64-bit

Furthermore, examining the overall Geek bench score for the current versions of virtualization software gives us the same ranking results in which Boot Camp™ > Parallels™ 8 > VMware Fusion™ 5 > Virtual Box 4.2. [8]

## 2.11 Cinebench

"Cinebench is a multi-platform benchmarking utility that is based on Maxon's Cinema 4D rendering software. It tests OpenGL graphics performance and multi- and single-CPU rendering capabilities." [8] Similar to the tests above testing GPU, Virtual Box could not operate the OpenGL portion of the Cinebench test. It did manage to run the rendering tests, and was given; in the chart a score of zero for its OpenGL results, Fig. 8. [8]

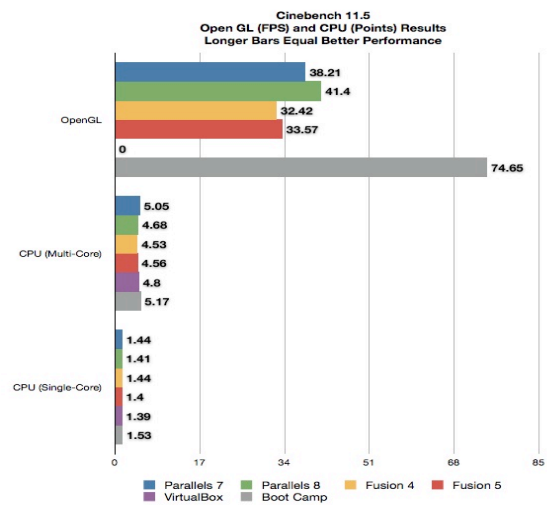

Fig. 8: Cinebench 11.5

Parallels™ 8 through rigorous testing scored as the highest of all the virtualization applications on the OpenGl test with a grand total of 41.4 frames per second. However, all virtualization software fell well in range simply at 74.65 frames per second behind native performance. [8] In terms of rendering performance however, it was a much tighter. Boot Camp™ tests were limited to a mere four processors, but the performance was still surprisingly better than virtualized options. Parallels™ 7 held a slender lead among the virtualization software and VirtualBox compared to its commercial rivals scored well, taking the second place position in the multi-core test. [8]

## 2.12 Just Cause 2

In 2010, Eidos released" Just Cause 2" as an open-world action game which supports DirectX 10. Therefore, like the 3DMark Vantage test, Boot Camp™ and Parallels™ 8 were only tested via its experimental DirectX 10 mode. Four resolutions were tested on the "Desert Sunrise" built-in benchmark, Fig. 9." [8]

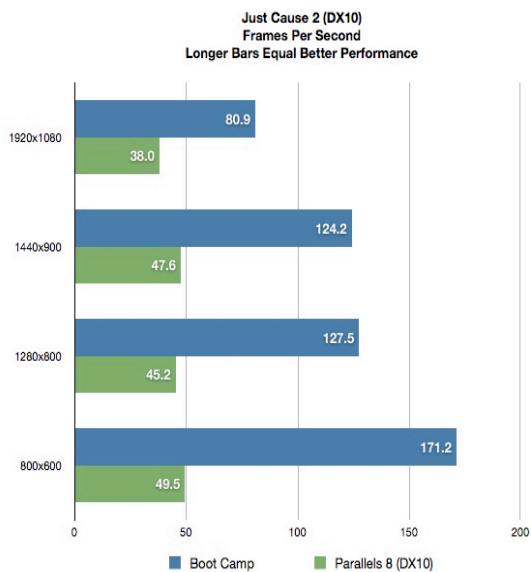

Fig. 9: Just Cause 2(DX10)

Parallels™ 8 offers operational frame rates, but is lacking in innate performance. Some issues appears also with Parallels'™ DirectX 10 driver, as resolutions between 1440x900 and 800x600 scored roughly the same. As advised, gamers should not yet plan to enjoy DirectX 10 gaming for this generation of virtual software. [8]

## 2.13 Crysis

Crytek in 2007 released "Crysis", which was once the pinnacle measure of a gaming PC's performance, testing even the most powerful machines. Expensive and capable machines were seemingly rendered obsolete. As a sign of how far technology has progressed in the past five years, Crysis is playable in a virtualized environment on a Mac. Competitive gamers who demand at least 60 frames per second will be disappointed, but more casual gamers who want to see if their Mac "can play Crysis," will be able to enjoy the game. [8] Implementing the Crysis a standard quality test was run in DirectX 9 mode at three different resolutions, Fig. 10. [8]

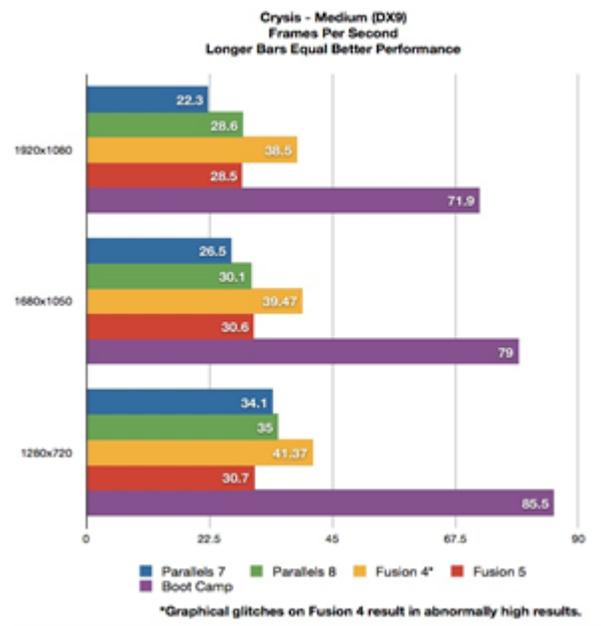

Fig. 10: Crysis

A huge and highly coerce performance surprisingly came from Boot Camp™ buts this is not to take away from the close second which was VMware Fusion™ 4. As stated below the chart, however, there were noticeable graphical glitches during the test. It was by all means still playable, but considering the glitches, lag time, and the performance of the other applications, VMware Fusion™ 4's results do not carry much weight. [8] Surprisingly, results were nearly identical with the exception of the 1280x720 resolution between Parallels™ 8 and VMware Fusion™ 5. Moreover, where thoughts that VMware Fusion™ will hit a virtualized processing limit, it scores the same as 1680x1050. [8]

2.14 Mafia II

Mafia II, a video game released in 2010 by the company known as 2K Games. It is centered on a third-person open-world adventure game. It runs in DirectX9 but would not run properly on VMware Fusion™ 4 or 5, either crashing before completing the benchmark or freezing and stuttering throughout the run. Due to these complications with these systems, they were excluded from the charts, Fig. 11. [8]

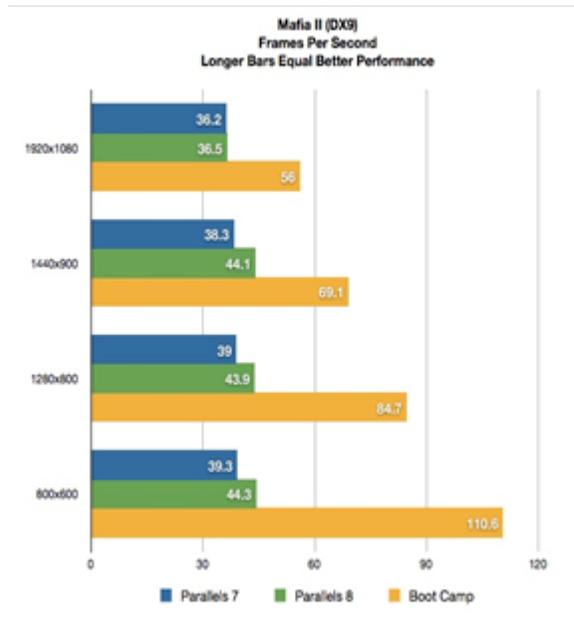

Fig. 11: Mafia II (DX9)

Boot Camp™ delivers noticeably enhanced performance; however, with that being said, the game does run at playable frame rates in Parallels™. Parallels™ 8 increased the frame rates by about 12 to 15 percent over Parallels™ 7, although there is a bottleneck in the way Parallels™ virtualizes the GPU, as the frame rates were approximately consistent across the bottom three resolutions. [8]

2.15 Boot Times

A clean installation of Windows 7 was used to test our boot times; five cold boots were run for each application recording time with a stopwatch. Then data was compiled we compiled and averaged out with the times and recorded the times rounding to the nearest second. [8] Timing was started when the power or start buttons were clicked to launch the virtual machine from within each application, and it ended when all items in the Windows system tray were loaded, Fig. 12. [8]

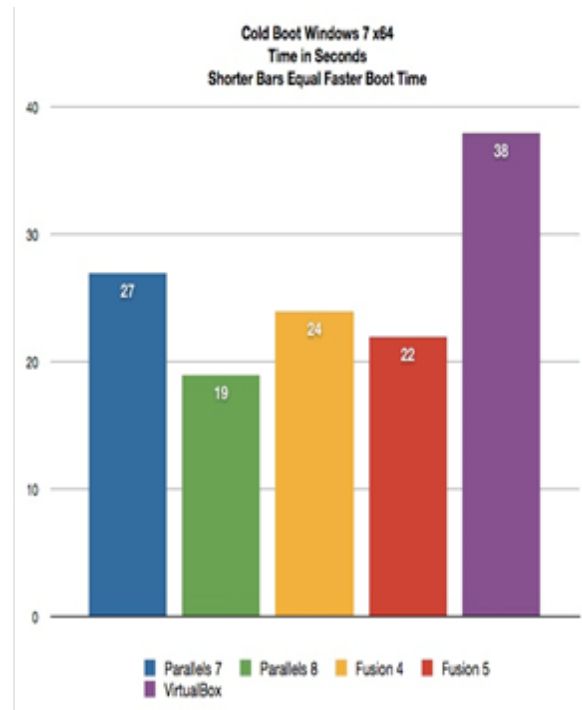

Fig. 12: Cold Boot Window 7

Through testing of boot times, the results showed that Parallels™ 8 was the fastest boot time at 19 seconds, in close second was VMware Fusion™ 5 at 22 seconds. So Parallel's™ time is almost a 42 percent advantage. [8] Virtual Box required the longest boot up time, at around roughly 38 seconds, which is twice as long as Parallels™ 8 and VMware Fusion™ 5. As long as users plan to keep a single virtual machine operating for a while, however, 20 additional seconds of boot time should not severely affect a user's experience. [8]

3. Benchmark Analysis

According to the industry observer website *MacTech*, two of the most popular virtualization products for the MacOS are Parallel™ Desktop and VMware Fusion™ 5. The MacTech article included an in-depth benchmarking analysis of the most recent versions of these two products. MacTech put the two apps to the test. The testing includes launching various OSes, performance of the applications, 3D graphic tests, and retina support. As a conclusion, Parallels™ did better on the majority of the tests and on 62% of the 3D graphics tests:
"In the vast majority of our overall tests, Parallels™ Desktop 8 won. Again, if count up the general tests (including the top 3D graphics scores), Parallels™ won 56% of the tests by 10% or more." [9] and "If include all the tests where Parallels™ was at least 5% faster, as well as the balance of the 3DMark06 graphics tests, Parallels™ increased the lead further." [9]

Fig. 13 presents comparative data on the MacOS X performance and the relative guest performance. The results show interesting information, that the processor integer performance has no difference between Lion and Mountain Lion, virtualized by VMware Fusion™ or Parallels™. However, the memory performance and memory bandwidth performance in parallels™ is better than VMware Fusion™ on both operating systems, Lion and Mountain Lion.

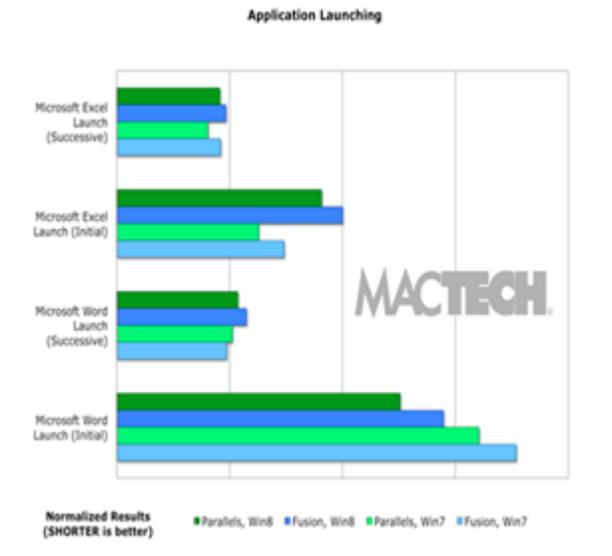

Fig. 14. Application Launching

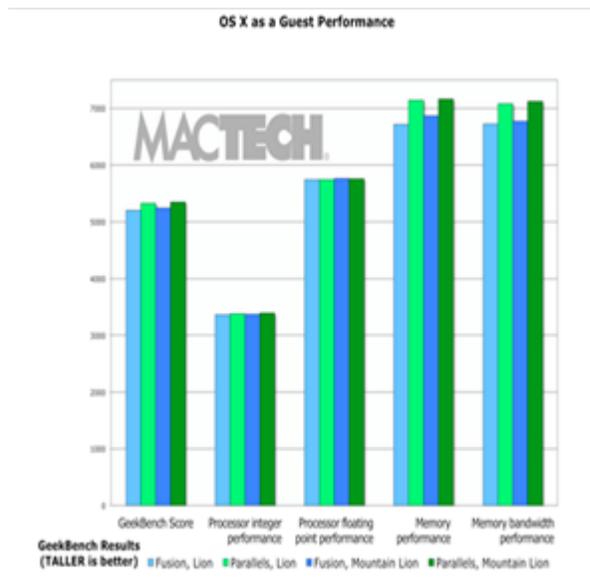

Fig. 13. OS X as Guest Performance

Fig. 14 presents the relative application launching. According to MachTech, Windows 7 on parallels™ launchs Microsoft Excel faster than Windows 8. Also, it's faster than launchs it on VMwareFusion™ with Windows 7 and Windows 8. In the other hand, Microsoft Word initial launch on parralels™ with Windows 8 is fasert than windows 7 and it is faster than VMware Fusion™ in becth opreaubg stsrem as well.

The conclouded information from Fig. 15 is fastest boot time in Windows 7 on Parallels™. Both Parallels™ and VMware Fusion™ boot Windows 7 faster than Windows 8. Parallels™ Shut down Windows 8 faster than Windows 7 but VMware Fusion™ shut down Windows 7 faster than Windows 8. The fastest shut down time is parallels™ with Windows 8. The case of compressing 1 GB file Windows 7 on VMware Fusion™ did the best job.

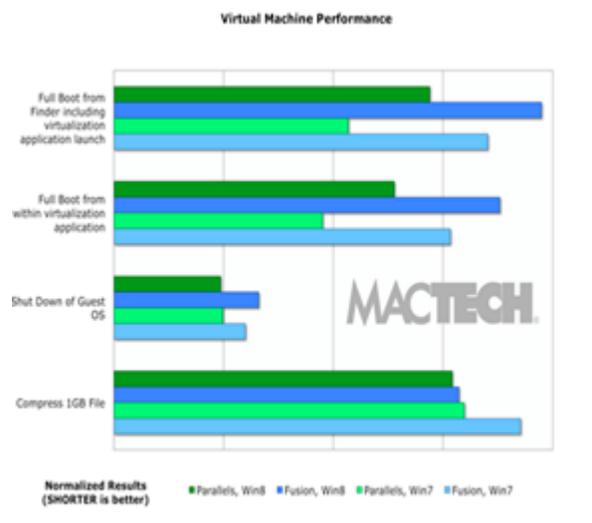

Fig. 15: Virtual Machine Performance

The graphical comparison of Parallels™ Desktop and VMware™: In this section, full comparison in different areas been made. The comparisons include installing Windows on Mac, running Windows applications on Mac, making Windows safer on Mac, and advance tools. Table 1 is about installing Windows on Mac. The similarities are ability of automatic setup; import physical Windows PC, import Boot Camp™ partition, and Windows support. [10]

Table 1: Install Windows on Mac

| Feature | VMware Fusion™ | Parallels Desktop 8™ |
|---|---|---|
| Easy Install with Automatic Windows Setup | Yes | Yes |
| Run off existing Boot Camp partition | Windows XP (32-bit), Windows Vista (32-bit and 64-bit), Windows 7 (32-bit and 64-bit) | Windows XP (32-bit), Windows Vista (32-bit and 64-bit), Windows 7 (32-bit and 64-bit) |
| Allows suspending of VM running off the Boot Camp partition | Yes | Yes |
| Import Physical Windows PC to VM | Yes | Yes |
| Import Boot Camp partition to VM | Yes | Yes |
| Full Support for Windows 7 | Yes | Yes |
| Full Support for Windows 8 | Yes | Yes |
| Import Third Party VMs (Parallels, VMware, VirtualPC for Mac) | Yes | Yes |

Table 2 is running windows applications on Mac. The similarities are running Windows apps like Mac apps (Unity/Coherence), application menu available to launch Windows apps any time, Windows start menu in menu, and drag and drop files between Windows and Mac. The differences are running Windows apps in Retina resolution (Unity/Coherence), Windows application folder in menu bar, and enable Apple remote to work with windows applications. [10]

Table 2: Running Windows Applications on Mac

| Feature | VMware Fusion™ | Parallels Desktop 8™ |
|---|---|---|
| Run Windows apps in Retina resolution (Unity/Coherence) | No | Yes |
| Support Mountain Lion Dictation in Windows 7/8 | No | Yes |
| Run Windows apps like Mac apps (Unity/Coherence) | Yes | Yes |
| Windows Start Menu in Dock | Yes | Yes |
| Windows Start Menu in menu | Yes | Yes |
| Windows application folder in Dock | No | Yes |
| Windows application folder in menu bar | Yes | No |
| Always On Application Menu available to launch Windows apps at any time | Yes | Yes |
| Quit Individual Window applications | Yes | Yes |
| Use Command ` to switch between open windows in a Windows app | Yes | Yes |
| Assign Windows applications to a Mac OS X Space | Yes | Doesn't keep windows from app together |
| Shared Folders to access Mac files/folders from Windows | Yes | Yes |
| Access Windows tray icons in Unity/Coherence | Yes | Yes |
| Arrow icon in the Mac menu bar for customizing Windows tray icons in Coherence. | No | Yes |
| Progress for downloads and other operations is displayed on the Windows 7 applications icons in the Dock. | No | Yes |
| Windows 7 Jump Lists are supported for Windows applications in the Dock: view your recent documents by right-clicking the application icon in the Dock. | No | Yes |
| Grouping of all windows of a single Windows application under the same application icon in the Dock. | No | Yes |
| Active screen corners | No | Yes |
| Drag and Drop Files Between Windows / Mac | Yes | Yes |
| Windows applications can be set as the default applications for handling CDs and DVDs inserted into the Mac. | No | Yes |
| Enable Apple Gestures to work with Windows applications | No | Yes |
| Enable Apple Magic Mouse Gesture support with Windows applications | No | Yes |
| Enable Apple Remote to work with Windows applications | No | Yes |
| Launch Mac applications from any Windows file (Shared Applications/SmartSelect) | Yes | Yes |
| Automatically mount storage and network devices to guest OS | Yes | Yes |

Table 3 is making Windows safer on Mac. The similarities are single and multiple snapshot support, and auto protect automatic snapshots. The differences are Mac OS Anti-Virus included, isolated Virtual Machines, and automatically revert VM to start state upon termination. [10][11].

Table 3: Making Windows safer on Mac

| Feature | VMware Fusion™ | Parallels Desktop 8™ |
|---|---|---|
| Single Snapshot support | Yes | Yes |
| Multiple Snapshot support | Yes | Yes |
| AutoProtect Automatic Snapshots | Yes | Yes |
| TimeMachine backups can be synced with SmartGuard snapshots, to reduce the space required for backups. | No | Yes |
| Automatically revert VM to start state upon termination | No | Yes |
| Virtual machine encryption with AES algorithm for better security of your data (empowered by AES-NI hardware support on i5 and i7 CPUs) | No | Yes |
| Mac OS Parental Controls are automatically applied to the virtual machine for managing children's computer usage | No | Yes |
| Windows AntiVirus & AntiSpyware Included | 12-month subscription to McAfee VirusScan Plus. User Prompted to Install | 1-month subscription to Norton Internet Security or Kaspersky Internet Security. User Prompted to Install |
| Mac OS AntiVirus Included | No | 1-month subscription to Norton Internet Security for Mac or Kaspersky Security for Mac. User Prompted to Install |
| Lock down application and virtual machine settings to prevent changes | No | Yes |
| Isolated Virtual Machines | No | Yes |

Table 4 is advance tools. The similarities are resizing Virtual disks; control VMs with scripting option, and network (PXE) Boot. The differences is supporting switch graphics card on Mac for longer battery live. [10]

Table 4: Advance Tools

| Feature | VMware Fusion™ | Parallels Desktop 8™ |
|---|---|---|
| Virtual Disk Management | Yes, Integrated in Settings Editor | Yes, Standalone tool |
| Resize Virtual Disks | Yes | Yes |
| Advanced Network Management | Requires modifying networking scripts | Yes |
| Control VMs with Scripting option | Yes | Yes |
| Supports Intel VT-x hardware virtualization engine | Yes | Yes |
| Network (PXE) Boot | Yes | Yes |
| Support switching graphics card on Mac Book Pro for longer battery live | No | Yes |

## 4. Conclusion

To conclude, there should be no surprises in the results depending on the recent tests. Operators who require or seek the best performance should without any doubt use Boot Camp™. On the other hand, users who want a blend between performance and convenience should consider Parallels™ 8 or Fusion™ 5, although the Parallels™ OS has a slight advantage in performance. [8] With the end result showing Boot Camp™ as the best way to run Windows on Mac OS X, the original project proposal was proven. Boot Camp™ finished highest in most tests and showed time after time to beat the competition. Also, compared to the proposal Parallels™ it is one of the highest rated virtualizations. The speed and performance shown proved that the user would be satisfied with either way. In future versions of this experimentation, we could test future updates in each of these different options and the future operating system releases. Boot Camp™ and Parallels™ use the most ethical approaches to loading Windows. The user must use the manufacturer certified disc and have a product key in order to load Windows. If the user does not have these then the operating system will not fully load. Data access is no problem with Parallels™ or VMware Fusion™ 5 since they run side by side with OS X. Boot Camp™ data access would be more challenging since a re-boot is necessary every time. The benefit of using any of these systems outweighs the problems because of the flexibility of using Windows with Mac OS X. As proven earlier Boot Camp™ does this the best because of the performance it possesses. An organization would benefit greatly from Boot Camp™ because they can use what programs would be needed in any situation.

**Ayman NOOR,** earned a Master of Science in Computer and Information Science from Gannon University, Pennsylvania, U.S.A in 2013 and a Bachelor in Computer Science from the College o Computer Science and Engineering from Taibah University, Madinah, Saudi Arabia in 2006. Mr. NOOR is currently a Lecturer at the Department of Computer Science, Taibah University.